\documentclass[12pt]{article}
\usepackage[dvips]{graphicx}
\usepackage{epsfig}
\usepackage{float}
\usepackage{cite, comment}
\usepackage{latexsym,amsmath,amsfonts,amssymb}
\usepackage{mathtools} 
\usepackage{subcaption}
\usepackage{slashed} 
\usepackage{blkarray}
\usepackage[all]{xy} 
\usepackage[makeroom]{cancel}
\usepackage[latin1]{inputenc}
\usepackage[american]{babel}
\usepackage{bbm}
\usepackage{color}
\usepackage[unicode]{hyperref}
\def\6{\partial}

\newcommand{\be}{\begin{equation}}
\newcommand{\ee}{\end{equation}}
\newcommand{\beq}{\begin{equation}}
\newcommand{\eeq}{\end{equation}}
\newcommand{\bea}{\begin{eqnarray}}
\newcommand{\eea}{\end{eqnarray}}

\newcommand{\ba}{\begin{eqnarray}}
\newcommand{\ea}{\end{eqnarray}}

\newcommand{\beqs}{\begin{eqnarray}}
\newcommand{\eeqs}{\end{eqnarray}}
\newcommand{\bal}{\begin{aligned}}
\newcommand{\eal}{\end{aligned}}

\def\lbldef#1#2{\expandafter\gdef\csname #1\endcsname {#2}}

\def\href#1#2{#2}

\newcommand{\ber}{\begin{eqnarray}}
\newcommand{\eer}{\end{eqnarray}}

\newcommand{\beqar}{\begin{eqnarray}}

\newcommand{\eeqar}{\end{eqnarray}}

\newcommand{\dsl}
   {\kern.06em\hbox{\raise.15ex\hbox{$/$}\kern-.56em\hbox{$\partial$}}}

\newcommand{\eeqarr}{\end{eqnarray}}
\newcommand{\ZZ}{{\rm \kern 0.275em Z \kern -0.92em Z}\;}

\def\CC{{\mathchoice
{\rm C\mkern-8mu\vrule height1.45ex depth-.05ex
width.05em\mkern9mu\kern-.05em}
{\rm C\mkern-8mu\vrule height1.45ex depth-.05ex
width.05em\mkern9mu\kern-.05em}
{\rm C\mkern-8mu\vrule height1ex depth-.07ex
width.035em\mkern9mu\kern-.035em}
{\rm C\mkern-8mu\vrule height.65ex depth-.1ex
width.025em\mkern8mu\kern-.025em}}}

\def\RR{{\rm I\kern-1.6pt {\rm R}}}

\def\ZZ{{\rm Z}\kern-3.8pt {\rm Z} \kern2pt}
\def\IB{\relax{\rm I\kern-.18em B}}
\def\ID{\relax{\rm I\kern-.18em D}}
\def\II{\relax{\rm I\kern-.18em I}}
\def\IP{\relax{\rm I\kern-.18em P}}

\newcommand{\bear}{\begin{eqnarray}}
\newcommand{\eear}{\end{eqnarray}}

\def\6{\partial}

\newfont{\namefont}{cmr10}
\newfont{\addfont}{cmti7 scaled 1440}
\newfont{\boldmathfont}{cmbx10}
\newfont{\headfontb}{cmbx10 scaled 1728}

\newcommand{\dd}{\textrm{d}}

\allowdisplaybreaks[1]

\numberwithin{equation}{section}

\begin{document}

\begin{titlepage}

\vfill
\begin{flushright}
\end{flushright}

\vfill

\begin{center}
   \baselineskip=16pt
   {\Large \bf  Cosmography and flat $\Lambda$CDM tensions at high redshift}
   \vskip 2cm
   Tao Yang$^{a}$, Aritra Banerjee$^{a}$ \& Eoin \'O Colg\'ain$^{a, b}$
          \vskip .6cm
             \begin{small}
               \textit{$^a$ Asia Pacific Center for Theoretical Physics, Postech, Pohang 37673, Korea}

               \vspace{3mm}
               
               \textit{$^b$ Department of Physics, Postech, Pohang 37673, Korea}
               
                \end{small}
\end{center}

\vfill \begin{center} \textbf{Abstract}\end{center} \begin{quote}
{Risaliti, Lusso \& collaborators have constructed a high-redshift Hubble diagram of supernovae (SNe), quasars (QSO) and gamma-ray bursts (GRB) that shows a ``$\sim 4 \, \sigma$ tension with the $\Lambda$CDM model" based on a log polynomial cosmographic expansion \cite{Risaliti:2018reu, Lusso:2019akb}. In this work, we demonstrate that the log polynomial expansion generically fails to recover flat $\Lambda$CDM beyond $z \sim 2$, thus undermining the $\sim 4 \, \sigma$ tension claim. Moreover, through direct fits of both the flat $\Lambda$CDM and the log polynomial model to the SNe+QSO+GRB dataset, we confirm that the flat $\Lambda$CDM model is preferred. Ultimately, we trace the tension to the QSO data and show that a best-fit of the flat $\Lambda$CDM model to the QSO data leads to a flat $\Lambda$CDM Universe with no dark energy within $1 \, \sigma$. This marks an irreconcilable tension between the Risaliti-Lusso QSOs and flat $\Lambda$CDM.}
\end{quote} \vfill

\end{titlepage}

\section{Introduction}
To first approximation the Universe is well described by the $\Lambda$CDM cosmological model built on the assumptions of a cosmological constant and cold dark matter. Recently, this harmony has been disturbed by conflicting determinations of the Hubble constant $H_0$. Indeed, the disagreement between local measurements \cite{Riess:2019cxk, Wong:2019kwg, Freedman:2019jwv} and determinations based on the Cosmic Microwave Background (CMB) by the Planck Collaboration \cite{Aghanim:2018eyx} is now in a $4 \sim 6 \, \sigma$ window \cite{Verde:2019ivm} \footnote{{See \cite{Dutta:2019pio} for recent analysis questioning this picture.}}. In the wake of Hubble tension/problem/crisis, other tensions have emerged \cite{Joudaki:2019pmv}, which point to potential differences in matter density $\Omega_m$. Since $\Omega_m$ is suppressed by factors of redshift $z$ relative to $H_0$ in the Hubble parameter $H(z)$, these tensions are redshift dependent and understandably less well established: the statistical significance is about $2 \sim 3 \, \sigma$.  

Against this backdrop claims of $\sim 4 \, \sigma$ deviations \cite{Risaliti:2018reu, Lusso:2019akb} from flat $\Lambda$CDM based on a high-redshift Hubble diagram of Type Ia supernovae (SNe), quasars (QSOs) and gamma-ray bursts (GRBs) are eye-catching. Recently, great strides have been made towards the use of QSOs as standard candles \cite{Risaliti:2015zla} and Risaliti \& Lusso have succeeded in producing a large catalogue of 1598 QSOs in the redshift range $0.04 < z < 5.1$ \cite{Risaliti:2018reu}. When combined with Pantheon Type Ia SNe \cite{Scolnic:2017caz} \footnote{{In the Pantheon dataset the various nuisance parameters have been combined into an overall error, thus one no longer has to fit the nuisance parameters.}} and an existing GRB compilation \cite{Demianski:2016zxi, Demianski:2016dsa}, this results in 2800 odd spectroscopically confirmed objects in the redshift range $0.01 < z < 6.67$. QSO and GRB datasets show considerable promise as cosmological probes at higher redshifts.

Our goal in this paper is to show that the Risaliti-Lusso QSOs are inconsistent with the flat $\Lambda$CDM model, but not because there is a $\sim 4 \, \sigma$ deviation, as claimed, but because a best-fit of the QSO data to flat $\Lambda$CDM allows no room for dark energy within $1 \, \sigma$. Thus, this is no longer a tension, but may be regarded as a glaring inconsistency. Obviously, our narrative is model dependent and the claims of \cite{Risaliti:2018reu, Lusso:2019akb} rest on a log polynomial expansion, which is claimed to be ``model independent". Thus, our first task is to identify shortcomings with the log polynomial expansion, so that we can prepare the reader for the eventual punchline, which we believe is a valid interpretation of the Risaliti-Lusso QSO data \cite{Risaliti:2018reu}.

{To begin, recall} that the papers \cite{Risaliti:2018reu, Lusso:2019akb} quote a $\sim 4 \, \sigma$ deviation from flat $\Lambda$CDM for matter densities in the range $0.1 \leq \Omega_m \leq 0.9$ \footnote{This is the content of Fig. 3 of \cite{Risaliti:2018reu} and Fig. 4 of \cite{Lusso:2019akb}.}, which is a strong claim. In section \ref{sec:scene} we show that the log polynomial expansion typically fails to recover flat $\Lambda$CDM behaviour beyond $z \sim 2$. Interestingly, the breakdown in the approximation is dependent on both matter density $\Omega_m$ and the (constant) equation of state (EoS) in the $w$CDM model, which contradicts the assumption of model independence. This breakdown in approximation undermines the analysis of \cite{Risaliti:2018reu, Lusso:2019akb} and leads to a comparison between a log polynomial fit over the redshift range $0 \lesssim z \lesssim 7$ and the flat $\Lambda$CDM model restricted below $z \lesssim 2$. \textit{A priori}, one cannot preclude deviations from flat $\Lambda$CDM that are coming from the breakdown in this approximation. {In fact, it is easy to find extreme examples \cite{Banerjee:2020bjq} that highlight the flaws of the log polynomial expansion. In appendix \ref{sec:cosmo}, we expand on the problems of high redshift cosmography, as implemented in \cite{Risaliti:2018reu, Lusso:2019akb}, by providing some examples.}

{In section \ref{sec:data} we fit both the log polynomial expansion and flat $\Lambda$CDM directly to the SNe+QSO+GRB dataset, thereby facilitating a like-for-like comparison. We recover the the best-fit parameters of Lusso et al. \cite{Lusso:2019akb} within $1 \, \sigma$ \footnote{The discrepancy may be due to the fact that we treat $H_0$ as a free parameter.} and confirm that a best-fit of flat $\Lambda$CDM using the same methodology leads to the matter density $\Omega_m = 0.369^{+0.015}_{-0.014}$, which we confirm is $3.5 \, \sigma$ discrepant with Planck \cite{Aghanim:2018eyx}. Note, this confirms the claimed tension with flat $\Lambda$CDM for $\Omega_m \approx 0.3$ , namely the the standard model, but appears to be out of sync with the claim of a $\sim 4 \, \sigma$ deviation from flat $\Lambda$CDM for any value of $\Omega_m$ (see  Fig. 3 \cite{Risaliti:2018reu}, Fig. 4 \cite{Lusso:2019akb}). Throughout this analysis, the intrinsic dispersion, a measure of the internal scatter in the QSO data plays an interesting role: it appears to inflate the errors to the point that any distinction between flat $\Lambda$CDM and the log polynomial model is lost, and remarkably, the flat $\Lambda$CDM model is preferred simply on the grounds that it has fewer parameters.} 

In the final section \ref{sec:tension}, we isolate the QSO data, which is clearly the source of the tension and perform direct fits of the QSO data to the flat $\Lambda$CDM model. We find that the best-fit matter density is within $1 \, \sigma$ of $\Omega_m =1$, namely a flat $\Lambda$CDM Universe with no dark energy. So, while we believe the deviations from the flat $\Lambda$CDM model observed by \cite{Risaliti:2018reu, Lusso:2019akb} have been impacted by a breakdown in the log polynomial expansion (see also \cite{Banerjee:2020bjq}), there may still be a \textit{bona fide} inconsistency with the flat $\Lambda$CDM model: \textit{the QSO data appears inconsistent with dark energy within the flat $\Lambda$CDM model}. Given the fact that dark energy is at this stage well established, regardless of whether one considers SNe \cite{Perlmutter:1998np, Riess:1998cb}, CMB \cite{Aghanim:2018eyx} or BAO \cite{Eisenstein:2005su}, this appears to imply that the Risaliti-Lusso QSOs are inconsistent with flat $\Lambda$CDM \footnote{See \cite{Velten:2019vwo} for a model independent analysis, where it is also concluded that the Risaliti-Lusso QSO data shows no evidence for late-time accelerated expansion. It should be stressed that the authors of this work have not manifestly taken the intrinsic dispersion (an internal error in the QSO data) into account, so we expect this may weaken their conclusions.}. Note, this is an alternative and stronger conclusion than the original $\sim 4 \, \sigma$ deviation from flat $\Lambda$CDM  \cite{Risaliti:2018reu, Lusso:2019akb}. 

\section{Setting the scene}
\label{sec:scene}
To get a better handle on this $\sim 4 \sigma$ deviation from flat $\Lambda$CDM, it is instructive to look at the numbers of Lusso et al. \cite{Lusso:2019akb}. The basic idea of the study \cite{Risaliti:2018reu} is to expand the luminosity distance in powers of $\log_{10} (1+z)$: 
\begin{eqnarray}
\label{dl}
d_{L}(z) &=&   \frac{c \ln (10)}{H_0} \biggl[ \log_{10} (1+z) + a_2 \log_{10}^2(1+z) \nonumber \\ &+& a_3 \log_{10}^3 (1+z)+ a_4 \log_{10}^4(1+z) + \dots  \biggr], 
\end{eqnarray}
where $H_0, a_2, a_3$ and $a_4$ are free parameters. The leading term in this expansion is fixed by requiring $H(z = 0) = H_0$ and this can be confirmed through the following identity for the Hubble parameter $H(z)$, 
\begin{equation}
\label{identity}
H(z) = \left[ \frac{d}{d z} \left( \frac{d_L(z)}{c (1+z)} \right) \right]^{-1}. 
\end{equation}
Since flat $\Lambda$CDM has only two parameters $(H_0, \Omega_m)$ at late times, to identify the remaining parameters $a_i$ in terms of the standard model, one expands the above luminosity distance and  the flat $\Lambda$CDM luminosity distance around $z = 0$. Making a comparison term by term, one identifies the following relations \cite{Lusso:2019akb}:
\begin{eqnarray}
\label{a2} a_2 &=& \ln (10) \left( \frac{3}{2} - \frac{3}{4} \Omega_m, \right),  \\
\label{a3} a_3 &=&  \ln^2 (10) \left( \frac{9}{8} \Omega_m^2 - 2 \Omega_m + \frac{7}{6} \right), \\
\label{a4} a_4 &=& \ln^3(10) \left(- \frac{135}{64} \Omega_m^3 + \frac{9}{2} \Omega_m^2 - \frac{47}{16} \Omega_m + \frac{5}{8} \right).  
\end{eqnarray}

Having discussed some preliminaries, let us return to the numbers. The best-fit value for the cosmographic model (see equation \ref{dl}) are reproduced from \cite{Lusso:2019akb} in Table 1. As is clear from the numbers, in particular $a_4$, there is some tension between the ``SNe, GRBs" dataset and the datasets involving QSOs.  However, it should be stressed here that the expansion parameters have no physical meaning and to make sense of the numbers, in particular in the context of flat $\Lambda$CDM, one can exploit equations (\ref{a2}) - (\ref{a4}) to recast the data in terms of matter density $\Omega_m$, which is the physical parameter. The result is shown in Table 2. \footnote{We do not have access to the MCMC chains of Lusso et al. \cite{Lusso:2019akb}, so here we have simply estimated the errors in the conventional fashion, $\Delta a_i = | \frac{\dd a_i}{\dd \Omega_m} | \Delta \Omega_m$. Later, we provide our own MCMC results. } .

\begin{table}[htb]
\centering
\begin{tabular}{llll}
\hline
\hline
\rule{0pt}{3ex} sample & $a_2$ & $a_3$ & $a_4$ \\
\hline 
\rule{0pt}{3ex}  SNe, quasars, GRBs & $3.205^{+0.165}_{-0.162}$ & $3.564^{+0.916}_{-0.938}$ & $-2.510^{+1.510}_{-1.536}$  \\
\rule{0pt}{3ex}  SNe, quasars & $3.075^{+0.172}_{-0.169}$ & $4.466^{+1.013}_{-1.040}$ & $-3.716^{+1.922}_{-1.852}$ \\
\rule{0pt}{3ex}  SNe, GRBs & $3.304^{+0.186}_{-0.183}$ & $2.069^{+1.122}_{-1.252}$ & $2.571^{+2.631}_{-2.506}$ \\
\hline 
\end{tabular}
\caption{Best-fit values of $a_i$ reproduced from Table 2 \cite{Lusso:2019akb}.} 
\end{table} 

Now that we have rewritten the results of \cite{Lusso:2019akb} in terms of the physical variable in the flat $\Lambda$CDM model, we can make a number of comments. The first thing to note is that the last entry, i.e. ``SNe, GRBs", is for all extensive purposes consistent with the standard model, or alternatively, flat $\Lambda$CDM with $\Omega_m \approx 0.3$ \footnote{{Interestingly, the numbers here, which come from \cite{Lusso:2019akb}, contradict the statement in the abstract of the same paper: ``The completely independent high-redshift Hubble diagrams of quasars and GRBs are fully consistent with each other, strongly suggesting that the deviation from the standard model is not due to unknown systematic effects but to new physics".  To see this, note that the Pantheon dataset is consistent with flat $\Lambda$CDM with $\Omega_m \approx 0.3$, so the GRBs must also be consistent. See SNe, GRBs entry in Table 2. In appendix \ref{sec:cosmo} we independently check that the GRB data is consistent with flat $\Lambda$CDM.}}. Secondly, the $\Omega_m$ value inferred from $a_2$ is slightly low relative to Planck \cite{Aghanim:2018eyx}, but this is easily explained: the Pantheon SNe dataset has a preference for a low value of $\Omega_m$ at low redshift $z \lesssim 0.2$ \cite{Colgain:2019pck, mvp, Camarena:2019moy} and this is where $a_2$ is most relevant. Thirdly, $a_3$ is consistent with Planck, but this is no surprise, since the QSOs have been calibrated by the Pantheon dataset at redshifts $ z \lesssim 1.4$, and Pantheon is consistent with Planck \cite{Scolnic:2017caz}, so this is also expected.  Lastly, while the $\Omega_m$ value inferred from $a_3$ is consistent with Planck within 1 $\sigma$, it is clear that the $\Omega_m > 1$ values corresponding to the QSO data are inconsistent with flat $\Lambda$CDM. So, naively there is an obvious inconsistency with flat $\Lambda$CDM. 

That being said, this is an inaccurate review of the findings of \cite{Risaliti:2018reu, Lusso:2019akb}. In essence, what the authors do is translate the $\Omega_m$ parameter of flat $\Lambda$CDM in the range $0.1 \leq \Omega_m \leq 0.9$ into the $a_i$ parameter space via (\ref{a2}) - (\ref{a4}) and show that the line describing the flat $\Lambda$CDM model is $\sim 4 \, \sigma$ removed from the best-fit values of $a_i$ to a combined SNe+QSO(+GRB) dataset. This point deserves emphasis: regardless of the value of $\Omega_m$, there is a $\sim 4 \, \sigma$ deviation from the flat $\Lambda$CDM family of cosmologies, and not just the standard model. Note, this is not a blatant inconsistency, but a significant deviation. 

\begin{table}[htb]
\centering
\begin{tabular}{llll}
\hline
\hline
\rule{0pt}{3ex} sample & $\Omega_m (a_2)$ & $\Omega_m (a_3)$ & $\Omega_m (a_4)$ \\
\hline 
\rule{0pt}{3ex}  SNe, quasars, GRBs & $0.144^{+0.096}_{-0.094}$ & $0.297^{+0.130}_{-0.133}$ & $1.291^{+0.066}_{-0.067}$  \\
\rule{0pt}{3ex}  SNe, quasars & $0.219^{+0.100}_{-0.098}$ & $0.180^{+0.120}_{-0.123}$ & $1.339^{+0.071}_{-0.068}$ \\
\rule{0pt}{3ex}  SNe, GRBs & $0.087^{+0.108}_{-0.106}$ & $0.573^{+0.298}_{-0.332}$ & $0.193^{+0.150}_{-0.143}$ \\
\hline 
\end{tabular}
\caption{Best-fit values of $\Omega_m$ inferred from the various expansion parameters.} 
\end{table}

\subsection{Comment on log polynomial model} 
\label{sec:comment}
{Throughout \cite{Risaliti:2018reu, Lusso:2019akb}, the authors have employed cosmographic expansions that are claimed to be ``model independent" and eschew fitting the flat $\Lambda$CDM model directly. We believe this has led to considerable confusion to the point that the conclusions of \cite{Risaliti:2018reu, Lusso:2019akb} have been heavily impacted. That being said, there is a rational for using cosmographic expansions, since if executed correctly, one can not only confront a specific model with data, but a larger class of models. For example, Taylor expansion in $z$ at low redshift $z$ is suitably general that it captures a large class of models and only assumes analyticity, so it certainly has merit.} 

Before identifying the key shortcoming with the log polynomial expansion, let us first comment on what it gets right. Whenever one employs a truncation, it is important to make sure that the truncation makes sense and this will only be true provided higher order terms are suppressed by the expansion parameter in the range of interest. In other words, we require that 
\be
|a_n| > |a_{n+1} | \log_{10} (1+z). 
\ee
Now, since $\log_{10} (1+z) \lesssim 0.9 $ in the range of interest, one can confirm that all the values in Table 1 satisfy this condition, so the truncation makes sense and it is conceivable that the dropped terms are smaller.

{However, we can recall the analysis of ref. \cite{Cattoen:2007sk}, where it is noted that the scale factor $ a \equiv (1+z)^{-1}$ is singular at $z=-1$, thereby ensuring that the radius of convergence can be \textit{at most} $|z|=1$. Applying the same logic to  $\log_{10} ( 1 +z)$, or any power of $\log_{10} ( 1 +z)$, we see that it is also singular at $z=-1$ and that its radius of convergence is also at most $|z|=1$.  These conclusions follow from the Cauchy-Hadamard theorem \cite{CH} \footnote{ For any power series in one complex variable $z$, $f(z) = \sum_{n=0}^{\infty} c_n (z-a)^n$, $a, c_n \in \mathbb{C}$, then the radius $R$ of $f$ at the point $a$ is given by $1/R = \limsup\limits_{n \rightarrow \infty} (|c_n|^{1/n})$.}. Now, noting that redshift $z$ is an observable, which one typically expands around $z=0$ today, one can see that since (\ref{dl}) diverges at $z = -1$, then there must be some coefficient in the expansion that also diverges. This ensures that the radius of convergence of (\ref{dl}) is restricted to $|z| < 1$. This result from complex analysis implies that (\ref{dl}), no matter how many powers one considers, should not be trusted above $z \sim 1$. The purpose of the rest of the section is to present the consequences of this mathematics theorem in complex analysis in an accessible manner.}

\begin{figure}[H]
\centering
\begin{subfigure}{.5\textwidth}
  \centering
  \includegraphics[width=80mm]{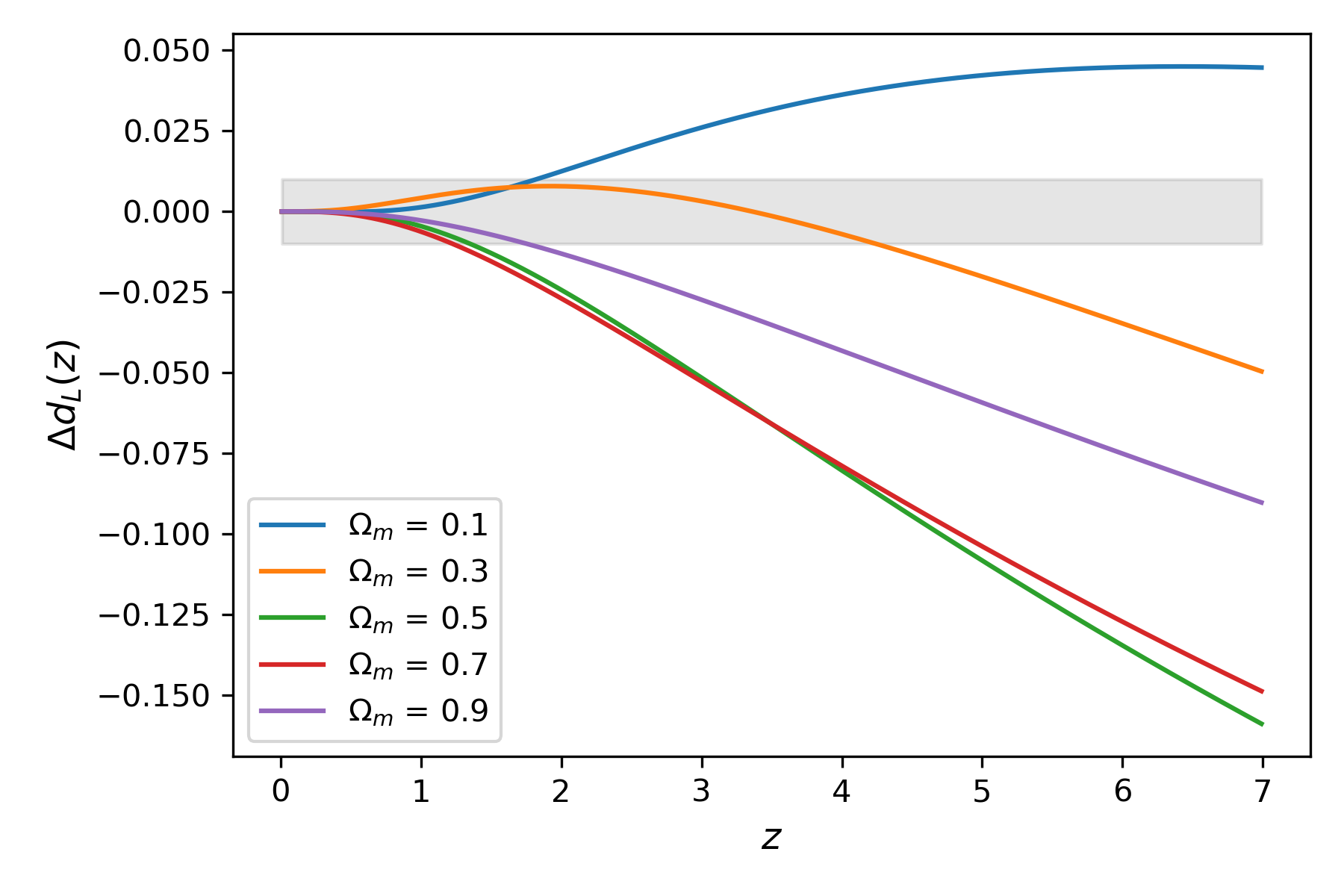}
  \caption{}
\end{subfigure}%
\begin{subfigure}{.5\textwidth}
  \centering
  \includegraphics[width=80mm]{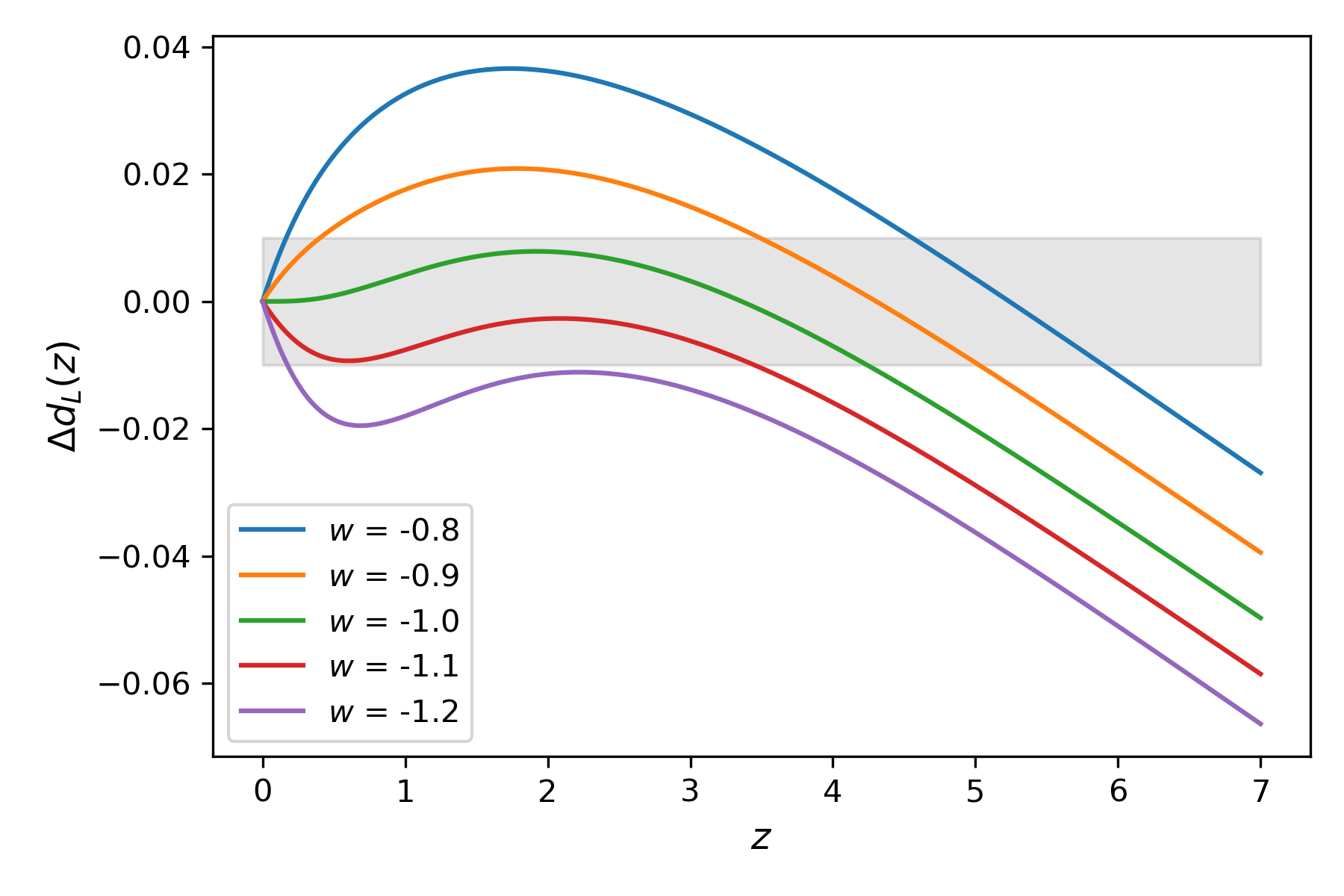}
  \caption{}
\end{subfigure}
\caption{\% difference between the fourth order log polynomial and flat $\Lambda$CDM based on different values of $\Omega_m$ in the left hand plot. The right hand plot presents the \% difference between the fourth order log polynomial and $w$CDM with $\Omega_m = 0.3$. In both plots we have shaded the region with $1$\% error or less.}
\label{approx_LCDM_wCDM}
\end{figure}

{We will now show that the} ``deviation from flat $\Lambda$CDM" conclusion needs to be taken with caution. It has a very important shortcoming, which ultimately undermines analysis based on the log polynomial expansion. In  Fig. \ref{approx_LCDM_wCDM} (a)  we illustrate the approximation inherent in the the fourth order log polynomial employed in \cite{Lusso:2019akb} (recently extended to fifth order in \cite{Lusso:2020pdb}) by comparing it to flat $\Lambda$CDM with different values of $\Omega_m$. Concretely, we plot the difference in the luminosity distance  
\begin{equation}
\Delta d_{L} (z) = \frac{d_{L}^{\textrm{poly}}(z) - d_{L}^{ \textrm{model}}(z)}{d_{L}^{ \textrm{model}}(z)}, 
\end{equation}
where it should be noted that $H_0$ drops out and we plot only odd values of $\Omega_m$ to avoid unnecessary clutter. There is no doubt that this is an interesting plot. Let us break it down. Evidently, in the redshift range $0 \lesssim z \lesssim 7$ the deviation from flat $\Lambda$CDM is smallest for $\Omega_m = 0.3$, where the approximation starts to deviate at the $1$\% level around $z \approx 4.3$. Note, for other values of $\Omega_m$, this deviation happens in the redshift range $1 \lesssim z \lesssim 2$. This essentially undermines the claim that the log polynomial expansion is ``model independent". Even within the flat $\Lambda$CDM family, the log polynomial expansion clearly approximates $\Omega_m = 0.3$, much better than the other cosmologies, so it is evidently biased towards certain models. One can check that the addition of an extra parameter in the log polynomial expansion \cite{Lusso:2020pdb} fails to change this conclusion \cite{Banerjee:2020bjq}. 

We can extend this analysis to the $w$CDM model with $\Omega_m = 0.3$ through a generalisation of (\ref{a2}) - (\ref{a4}) that includes the EoS $w$. The expressions are lengthy, so we omit them. The result is shown in Fig. \ref{approx_LCDM_wCDM} (b). It is clear from the plot that while $w = -1$ performs reasonably well over an extended redshift range, at higher redshifts the discrepancy between the log polynomial approximation to the luminosity distance and the actual luminosity distance exceeds $1$\% by a considerable margin. In essence, what this is saying is that if the underlying data is consistent with the standard model, then the log polynomial performs well, but if the data prefers a different value of $w$, then the fitting will be impacted by a breakdown in the approximation. 

So what have we learned? The fourth order log polynomial approximation works best for the standard model. In other words, if the data is close to $\Omega_m = 0.3$ within the flat $\Lambda$CDM model, then the approximation is under some control. However, this is merely a coincidence and it is clear that when one translates from physical parameters, such as $\Omega_m$ and $w$, to the unphysical $a_i$ parameters, the corresponding luminosity distance may deviate significantly from the original model. See \cite{Banerjee:2020bjq} for further comments on the breakdown of the approximation.

Ultimately, this means that all bets are off and there is no guarantee that the $\sim 4 \, \sigma$ deviations reported in \cite{Risaliti:2018reu, Lusso:2019akb} have not been impacted by the breakdown in the approximation to the luminosity distance. Let us dwell on this a bit further. In practice what Lusso et al. \cite{Risaliti:2018reu, Lusso:2019akb} are doing is translating flat $\Lambda$CDM into the log polynomial parameters $a_i$ using relations (\ref{a2}) - (\ref{a4}), which are \textit{only guaranteed to hold at low redshift}. This story is confirmed by Fig. \ref{approx_LCDM_wCDM}. Thus, the fallacy of \cite{Risaliti:2018reu, Lusso:2019akb} is that one is comparing $a_i$ based on flat $\Lambda$CDM at low redshift against a log polynomial expansion fitted over an extended redshift range $0 \lesssim z \lesssim 7$.  This is not a good comparison.

Given the difficulty accounting for this deviation in the analysis of \cite{Risaliti:2018reu, Lusso:2019akb}, it  is imperative to fit the flat $\Lambda$CDM model directly to the data to have any confidence in the claims. {In appendix \ref{sec:cosmo}, we collect some results on how the breakdown in approximations may lead to phantom tensions in high-redshift cosmography.} 

\section{Log polynomial versus $\Lambda$CDM} 
\label{sec:data}
In section \ref{sec:scene} we have shown that the log polynomial expansion shows a model bias, which contradicts the claim that it is model independent. More seriously, the luminosity distance as approximated by the log polynomial expansion can deviate significantly from the luminosity distance of the exact model. This is essentially because relations such as (\ref{a2}) - (\ref{a4}) are based on a Taylor expansion around $z=0$ and \textit{ a priori} do not have to hold at higher redshifts. This is evident from Fig. \ref{approx_LCDM_wCDM}, where the approximation typically breaks down at higher redshift. 

\begin{figure}[H]
   \centering
   \includegraphics[width=100mm]{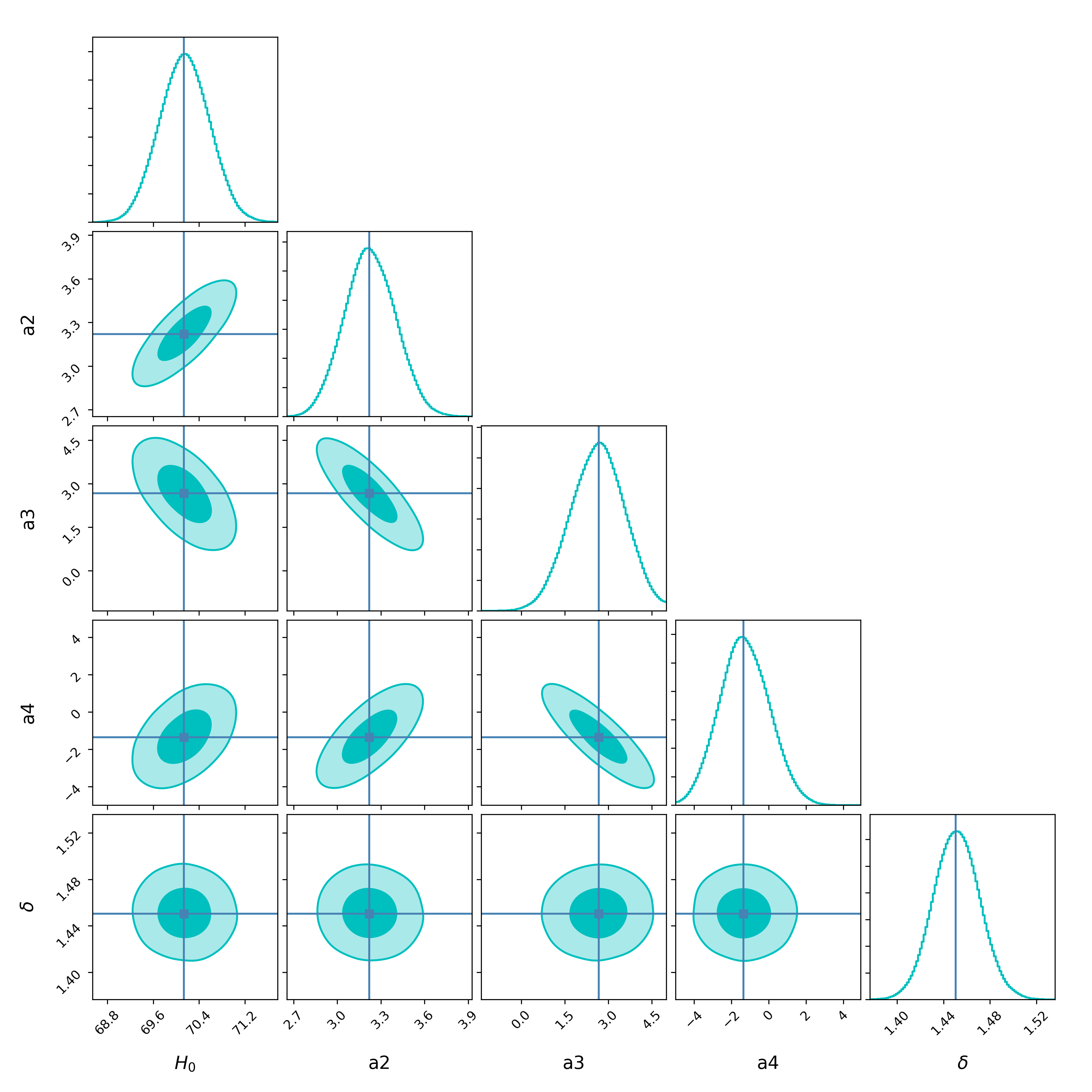}
   \caption{Best-fit values of the log polynomial model (\ref{dl}) to the overall SNe+QSO+GRB dataset. The lines denote the best-fit values from least squares fitting and they are consistent with the central values of the MCMC.}%
   \label{logpolymcmc}
    \end{figure}

Now, to get oriented, let us attempt to recover the results of \cite{Lusso:2019akb} in the original $a_i$ parameters. Taking into account the requisite shifts in the distance moduli (SNe $+19.63$, QSOs $+0$, GRBs $+0.54$), our best-fit values are shown in Fig. \ref{logpolymcmc} \footnote{{These offsets can be motivated on the grounds that the respective datasets, namely GRB, QSO and SNe, will agree on $H_0 \approx 70$ km/s/Mpc when restricted to $z \lesssim 1.4$. These are the offsets employed by Risaliti \& Lusso, which we take at face value. One could try to fit these values below $z \sim 1.4$ and propagate an error in the offsets, but this will only increase uncertainties. In particular, while it may weaken the $\sim 4 \sigma$ tension, it cannot change our punchline that a Universe with no dark energy is within $1 \, \sigma$ of best-fits of the flat $\Lambda$CDM model to the QSO data. On the contrary, it simply reinforces our narrative.}}. Note that $\delta$ is the intrinsic dispersion, which is a measure of the intrinsic scatter in the QSO data. Taking into account the fact that we are doing a global fit, e. g. \cite{Melia:2019nev}, versus fits with respect to different bins \cite{Risaliti:2018reu}, and the fact that we are working with the distance moduli and not the fluxes, one can confirm that our best-fit value of $\delta \approx 1.45$ is consistent with other determinations in the literature. Moreover, in the same plot, we show the best-fit values from least squares fitting and this provides a further consistency check.  

Our best-fit values and their corresponding $1\, \sigma$ confidence intervals are respectively $a_2 = 3.22^{+0.17}_{-0.17}, a_3 = 2.65^{+0.89}_{-0.93}$ and $a_4 = -1.32^{+1.33}_{-1.27}$. Interestingly, our best-fit values deviate from those quoted in \cite{Lusso:2019akb} (see Table 1) \footnote{Here, we treat $H_0$ as a free parameter, whereas in \cite{Lusso:2019akb} $H_0$ is fixed, so this is one possible explanation for the discrepancy.}, but nevertheless, they are consistent within $1 \, \sigma$. Thus, it is reasonable to assume that a similar analysis to \cite{Lusso:2019akb} in the parameter space ($a_2, a_3, a_4$) will also confirm a similarly large deviation from flat $\Lambda$CDM. Note, as explained earlier, this should be interpreted as a $\sim 4 \, \sigma$ deviation from the family of flat $\Lambda$CDM models based on any value of $\Omega_m$ in the range $0.1 \leq \Omega_m \leq 0.9$ \cite{Lusso:2019akb}. The reader should continually bear this in mind.  

\begin{figure}[H]
   \centering
   \includegraphics[width=100mm]{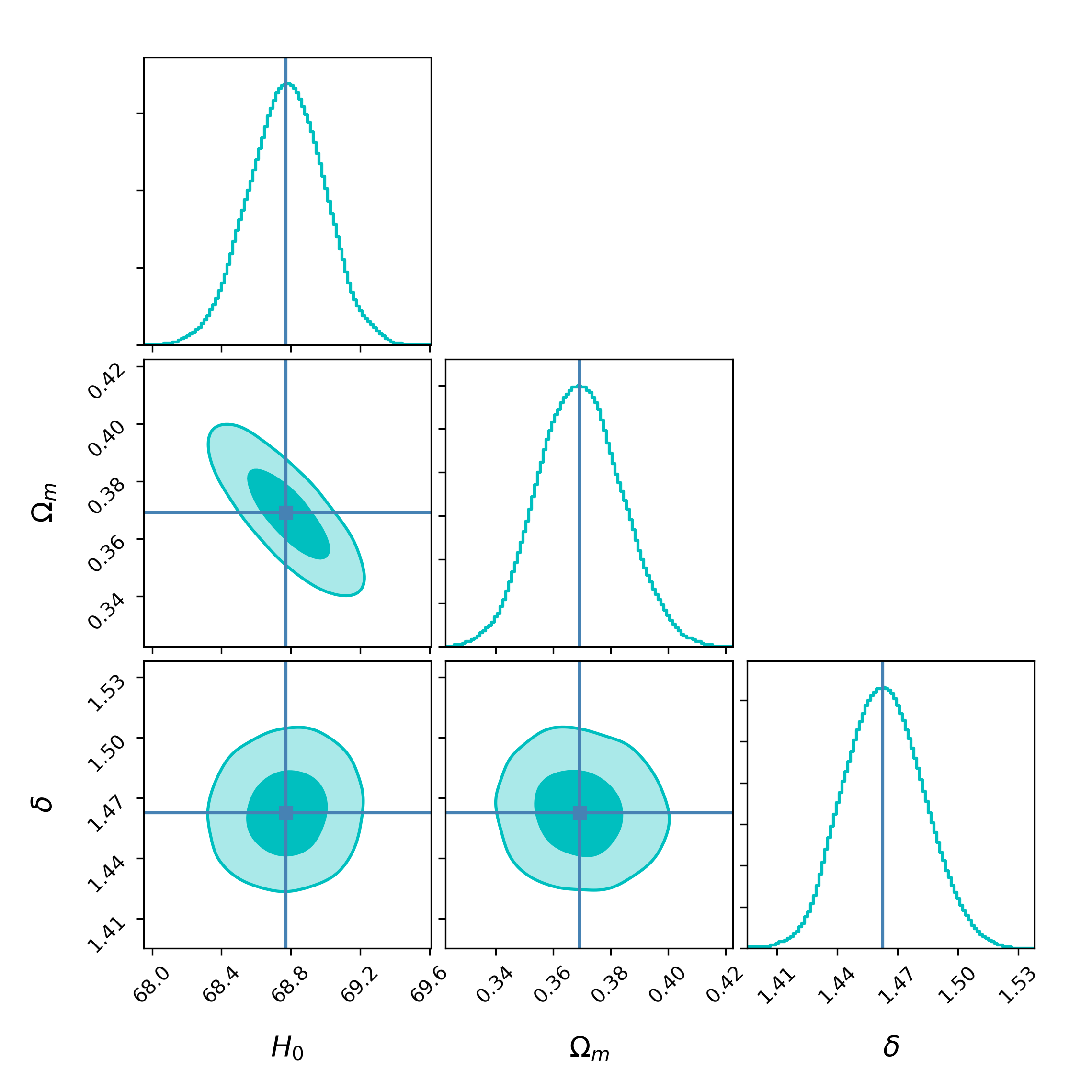}
   \caption{Best-fit values of flat $\Lambda$CDM to the overall SNe+QSO+GRB dataset. The lines denote the best-fit values from least squares fitting and they are consistent with the central values of the MCMC.}%
   \label{lcdm_SNeQSOGRB_mcmc}
    \end{figure}

Now, let us repeat the same exercise, but within the flat $\Lambda$CDM model. The result is shown in Fig. \ref{lcdm_SNeQSOGRB_mcmc} and for completeness we record our best-fit value of $\Omega_m$, $\Omega_m = 0.369^{+0.015}_{-0.014}$. It can easily be checked that this value is $\sim 3.5 \, \sigma$ discrepant with the Planck value, $\Omega_m = 0.315 \pm 0.007$ \cite{Aghanim:2018eyx}. So, there is a clear deviation from the standard model based on flat $\Lambda$CDM with Planck values. But given that Lusso et al. have claimed that their data is inconsistent with flat $\Lambda$CDM in the matter density range $0.1 \leq \Omega_m \leq 0.9$ \cite{Risaliti:2018reu, Lusso:2019akb}, it is a little bewildering that one can arrive at this conclusion. The only possibility is that the flat $\Lambda$CDM model fits the data considerably worse than the log polynomial and it can be discounted on those grounds.

So, let us look closely at the $\chi^2$ to see how the fourth order log polynomial and flat $\Lambda$CDM fit the data. {Before proceeding, a comment on the intrinsic dispersion $\delta$ is in order. It has been noted by Melia \cite{Melia:2019nev} (Table 1) that the reduced chi-square $\chi_{\nu}^2$ for a fit of the QSO data to any model is approximately unity, $\chi^2_{\nu} \approx 1$. In particular, there is little to distinguish the flat $\Lambda$CDM model and the third order log polynomial of \cite{Risaliti:2018reu}, and interestingly, Melia comes to the conclusion that flat $\Lambda$CDM is preferred by the data. There is an important difference here that Melia is not employing an external supernovae calibrator for the QSOs, but the results imply that the intrinsic dispersion dominates the uncertainties to the point that one can fit any model and $\chi^2_{\nu}$ will be close to unity. Thus, the observation becomes almost trivial. As is clear from Fig. \ref{real_mock_data} (a), the QSO data has a lot of internal scatter, so to overcome this,  one introduces an intrinsic dispersion parameter, which one proceeds to fit to the data. The effect of this term is to inflate all the statistical errors to the point that the data cannot distinguish differences in models, except in a trivial way. We believe this is a valid interpretation of the results of \cite{Melia:2019nev}.}

{Performing the analysis with the Risaliti-Lusso QSOs, which have been calibrated by supernovae \cite{Risaliti:2018reu}, we find that the $\chi^2$ values for the log polynomial expansion and flat $\Lambda$CDM are respectively $\chi^2 = 2791$ and $\chi^2 = 2788$ for $2806$ observations. Surprisingly, the log polynomial fits the data worse than flat $\Lambda$CDM, but the difference is marginal. The corresponding reduced chi-square values are $\chi^2_{\nu} = 0.996$ and $\chi^2_{\nu} = 0.995$, respectively. Now, recalling that the standard deviation of the reduced chi-square statistic is expected to be $\sqrt{\frac{2}{\nu}}$, we find that since $\nu \approx 2800$, the two values are equivalent within $1 \, \sigma$, in line with the findings of Melia \cite{Melia:2019nev}, modulo the fact that the QSO data has been calibrated differently. Moreover, both models with the inclusion of the intrinsic dispersion represent perfect fits to the data. Now, it is a simple exercise in counting parameters, and this favours the flat $\Lambda$CDM model.} 

\section{Direct fit of the QSO data}
\label{sec:tension} 
We have just confirmed that there is a tension between the SNe+QSO+GRB dataset and flat $\Lambda$CDM based on Planck values \cite{Aghanim:2018eyx} by fitting both the log polynomial expansion and the flat $\Lambda$CDM model directly. At this juncture, it makes sense to identify the source of the tension. Observe that we have confirmed through Fig. \ref{grbplot} that the GRB dataset is consistent with the standard model, so we can eliminate this possibility. 

Before turning our attention to the Risaliti-Lusso QSOs \cite{Risaliti:2018reu}, let us turn our focus to the Pantheon Type Ia SNe dataset \cite{Scolnic:2017caz}, where one has 1048 SNe in a range $0.01< z < 2.26$. Whether one employs least-square fitting or MCMC, it should come as no surprise that one recovers the $\Omega_m$ value from the fourth entry in Table 8 of \cite{Scolnic:2017caz}, thereby underscoring the fact that we have used the full Pantheon covariance matrix including both statistical and systematic uncertainties. More importantly, the best-fit value of $\Omega_m$ is consistent with the canonical flat $\Lambda$CDM value \footnote{However, see \cite{Colgain:2019pck, mvp, Camarena:2019moy} for a $2 \, \sigma$ discrepancy with flat $\Lambda$CDM at low redshift.}. The $+19.36$ shift in the distance moduli results in a best-fit value of $H_0 = 69.661^{+0.343}_{-0.344}\, \textrm{km s}^{-1} \textrm{ Mpc}^{-1}$, which is also within $1\, \sigma$ of the canonical value $H_0 = 70 \, \textrm{km s}^{-1} \textrm{ Mpc}^{-1}$. Therefore, as promised the Pantheon dataset is consistent with flat $\Lambda$CDM. 

\begin{figure}[ht]
   \centering
   \includegraphics[width=100mm]{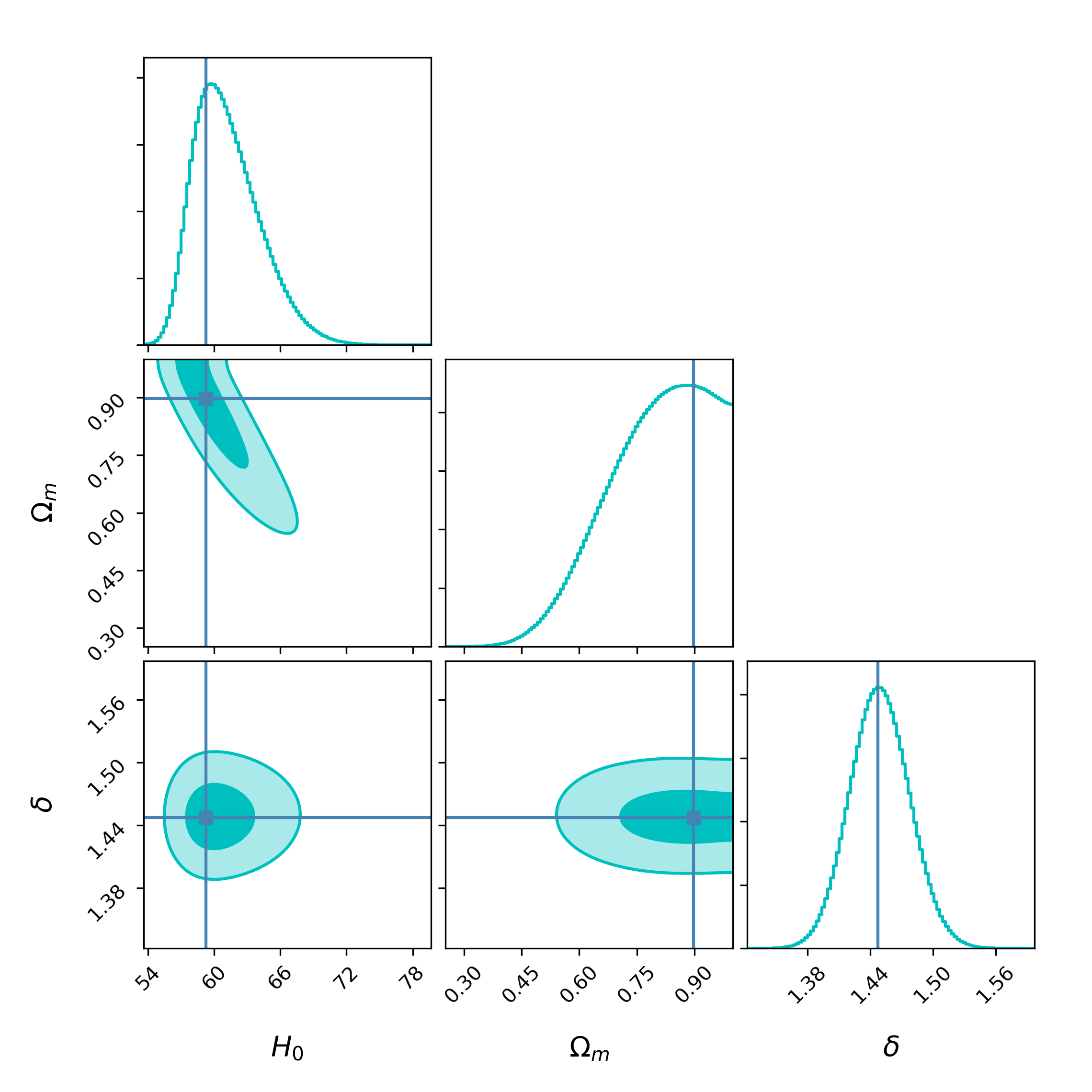}
   \caption{Best-fit values of flat $\Lambda$CDM to the QSO dataset.}%
   \label{qsoplot}
    \end{figure}

However, from the MCMC for the QSO data Fig. \ref{qsoplot}, it is clear that the best-fit value $\Omega_m \approx 0.9$ is in tension with the best-fit value of $\Lambda$CDM and $\Omega_m = 0.3$. There is also noticeable tension in $H_0$ with the SNe and GRB data. However, it is worth noting that a Universe without dark energy ($\Omega_m =1$) is within $1 \, \sigma$, and one has to address this contradiction before any $\sim 4 \, \sigma$ tension with flat $\Lambda$CDM can be taken seriously. Overall, this outcome may have been anticipated. The QSO data \cite{Risaliti:2018reu} suffers from considerable internal scatter and is of relatively poor quality compared to SNe, as is clear from Fig. \ref{real_mock_data} (a). As a result, a $\sim 4 \, \sigma$ deviation in matter density $\Omega_m$ can be expected to be far from the canonical value $\Omega_m = 0.3$.

\begin{figure}[ht]
   \centering
   \includegraphics[width=100mm]{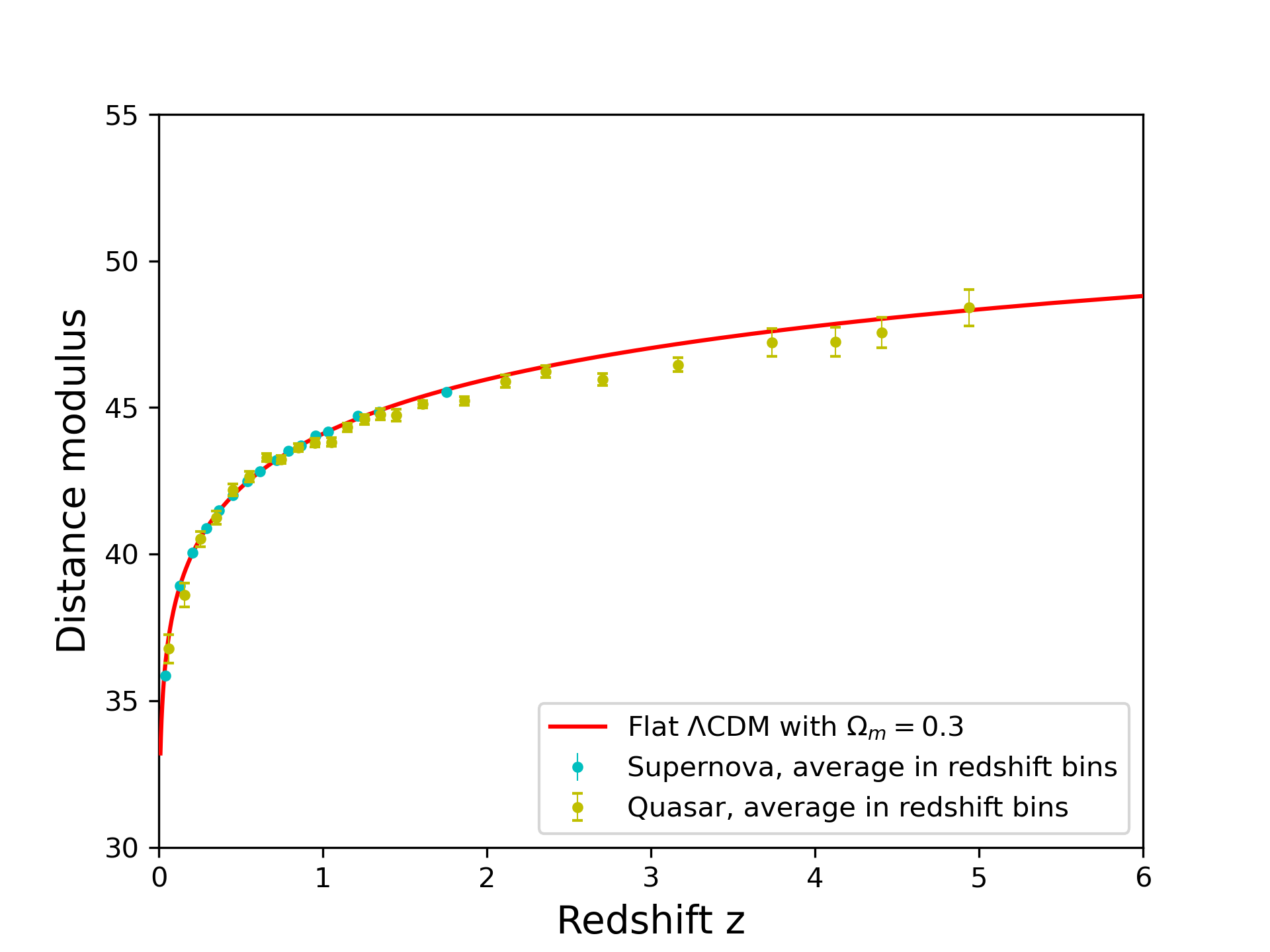}
   \caption{Comparison between the SNe and QSO binned data.}%
   \label{bindata} \end{figure}

Let us subject the Risaliti-Lusso data \cite{Risaliti:2018reu} to one more consistency check. To get a different perspective, it is a useful exercise to bin both the SNe and QSO data and compare. As is standard practice, we consider weighted means in each bin, e. g. \cite{Pinho:2018unz}: 
\begin{equation}
\bar{s}_i = \frac{\sum_{k}^{N_i} s_k (\sigma_k^s)^{-2}}{\sum_k^{N_i} (\sigma^s_k)^{-2}}, \quad \sigma_i^{\bar{s}} = \frac{1}{\sqrt{\sum_{k}^{N_i} (\sigma^s_k)^{-2}}} 
\end{equation}
where $s_k \equiv s (z_k)$ denotes the data value at each point $z_k$ with error $\sigma_k^s$ and $N_i$ is the number of data points in each bin $i$. $\bar{s}_i$ and $ \sigma_i^{\bar{s}} $ denote the new value and error for each bin. Note, in the binning process we adopt the weighted average value of $z_i$ for a given bin. The result of the binning procedure is illustrated in Fig. \ref{bindata}, where we have considered bins of length $\Delta z = 0.1$ at low redshift and bins of higher width where the data becomes sparser. From the plot it is clear that the SNe and QSO data follow flat $\Lambda$CDM with canonical values where they have been cross-correlated at $ z \lesssim 1.4$, thereby recovering the results of \cite{Risaliti:2018reu}, but at higher redshift the QSO data falls under the red curve and this ultimately explains the larger $\Omega_m$ value. This provides visual confirmation of the increase in matter density and drop off in dark energy density. 

Of course, one may complain that our analysis here is model dependent. However, see \cite{Velten:2019vwo} for an independent study of the deceleration parameter $q(z)$ in a model independent way where it is noted that the same QSO data provides little evidence for late-time acceleration, in line with our model dependent findings here. Since evidence for dark energy can be found in the Cosmic Microwave Background at even higher redshifts $z \approx 1100$, and BAO and Type Ia supernovae at lower redshifts, it is difficult to understand why QSOs calibrated by Type Ia SNe at intermediate redshifts see so little dark energy. Moreover, it is clear that if one does not calibrate QSOs with Type Ia supernovae, then one finds that the QSO data is consistent with flat $\Lambda$CDM \cite{Melia:2019nev, Khadka:2019njj}, so this just further deepens the puzzle.

\section{Discussion}
This work started with a desire to better understand the claim that a given SNe+QSO+GRB dataset is in ``$\sim 4 \, \sigma$ tension with the flat $\Lambda$CDM model" within a model independent framework \cite{Risaliti:2018reu, Lusso:2019akb}. Curiously, this statement holds for all matter densities in the range $0.1 \leq \Omega_m \leq 0.9$ \cite{Risaliti:2018reu, Lusso:2019akb} (Fig. 3 \& Fig. 4, respectively), so it is a bolder statement than a deviation from the standard model, i. e. flat $\Lambda$CDM with Planck values \cite{Aghanim:2018eyx}. Our work has identified a number of {shortcomings in} the analysis of \cite{Risaliti:2018reu, Lusso:2019akb}, which may be summarised as follows: 

\begin{itemize} 

\item Contrary to the claims, the log polynomial expansion pioneered in \cite{Risaliti:2018reu} cannot be model independent, since the approximation clearly varies with matter density $\Omega_m$ in the flat $\Lambda$CDM model and varies with EoS $w$ in the $w$CDM model. Given that the expansion parameters in the log polynomial expansion are unphysical, we believe that this negates the benefit in using this expansion. 

\item We have observed that the log polynomial luminosity distances inferred from flat $\Lambda$CDM typically deviate from the exact expression beyond $z \sim 2$. In practice, this means that the authors of \cite{Risaliti:2018reu, Lusso:2019akb} are constraining flat $\Lambda$CDM models to $z \lesssim 2$ when they translate from the physical parameter $\Omega_m$ to $a_i$ through (\ref{a2})- (\ref{a4}). These values are then compared to a best-fit log polynomial expansion over an extended redshift range $0 \lesssim z \lesssim 7$. One is implicitly comparing models in different redshift ranges. Note that the data below $z \lesssim 2$ is by construction consistent with the Pantheon dataset, but clearly deviates at higher redshift (see Fig. \ref{bindata}). 

\item {We have confirmed that one can fit the flat $\Lambda$CDM model directly to the SNe+QSO+GRB dataset and it is preferred over the log polynomial expansion \cite{Risaliti:2018reu, Lusso:2019akb}. Our findings here are in line with Melia \cite{Melia:2019nev}, and interestingly, we find that the intrinsic dispersion is so large that it reduces the question of which model is preferred to simply counting free parameters. It is not clear if the QSO data can distinguish models except in the most trivial manner.}
 
\item Evidently, \cite{Risaliti:2018reu, Lusso:2019akb} have implicitly combined datasets of different quality that have a preference for different cosmological parameters (including $H_0$). This can be confirmed by the jump in $\chi^2_{\nu}$ and shift in parameters when the QSO data is added. 
\end{itemize} 

Hopefully it is clear to the reader that the log polynomial analysis of Risaliti-Lusso obfuscates matters: there are certainly phantom deviations that are impacting the results and have led to misleading conclusions, most noticeably, the $\sim 4 \, \sigma$ deviation from flat $\Lambda$CDM for matter densities in the range $0.1 \leq \Omega_m \leq 0.9$ \cite{Risaliti:2018reu, Lusso:2019akb}. Ultimately, this makes the rational for cosmography at high-redshift unclear, since one has to work very hard to convince the reader that the expansion is not impacting results. It is simpler and more transparent to just do the obvious and fit models directly to the data.

While the expansion employed in \cite{Risaliti:2018reu, Lusso:2019akb} may have done the authors a disservice in negatively impacting the conclusions, there appears to be some truth to the tension claims. In particular,  our best-fit value for $\Omega_m = 0.369^{+0.015}_{-0.014}$ is discrepant from the Planck value at $3.5 \, \sigma$, thus confirming that there is a real tension with the standard model, namely flat $\Lambda$CDM with $\Omega_m \approx 0.3$. We traced this tension to the QSO data and confirmed that the best-fit value of $\Omega_m$ to the QSO data is consistent with a flat $\Lambda$CDM Universe with no dark energy. This marks an irreconcilable inconsistency between the Risaliti-Lusso QSOs and flat $\Lambda$CDM, which replaces the ``$\sim 4 \sigma$ tension from the $\Lambda$CDM model" claim. 

Our work here provides an alternative interpretation for the Risaliti-Lusso QSOs, which have been calibrated by Type Ia supernovae \cite{Risaliti:2018reu, Lusso:2019akb}, and highlights the seriousness of the conflict with dark energy within the flat $\Lambda$CDM model. However, given the current quality of the QSO data, it may be premature to conclude that the flat $\Lambda$CDM model is wrong. Nevertheless, going forward it will be interesting to see if future QSO datasets show similar tension with dark energy within flat $\Lambda$CDM, and if so, Lusso, Risaliti \& collaborators will deserve due credit if they can falsify the flat $\Lambda$CDM model. It is worth stressing again that a calibration exists whereby the QSOs are consistent with the standard model \cite{Melia:2019nev, Khadka:2019njj}, so it is possible that calibration is impacting the results \footnote{Interestingly, it has been noted recently that when one restricts QSOs to the redshift range $0.11 < z < 4.13$, the combined SNe+QSO dataset becomes consistent with the standard model \cite{Hu:2020mzd}.}. In the big picture, QSOs offer considerable promise to unlock high-redshift cosmology and it is imperative that the methodology is improved and the assumptions, for example the calibration, are tested further in a transparent manner.

\section*{Acknowledgements}
We thank Elisabeta Lusso \& Guido Risaliti for kindly sharing and explaining their Hubble diagram data. We thank Stephen Appleby, Eleonora Di Valentino, Lavinia Heisenberg, Fulvio Melia, Adam Riess, Misao Sasaki, Arman Shafieloo, Shahin Sheikh-Jabbari, Hermano Velten and Kenneth Wong for related discussions. We thank anonymous referees at Nature Astronomy, Astronomy \& Astrophysics and JCAP for encouraging us to dig deeper. This work was supported in part by the Korea Ministry of Science, ICT \& Future Planning, Gyeongsangbuk-do and Pohang City.

\appendix 

\section{Limitations of high-redshift cosmography}
\label{sec:cosmo}
In this section we make some self-contained comments on high-redshift cosmography. Even in its traditional form as a Taylor expansion around $z=0$ \cite{Visser:2003vq}, great care should be taken with cosmography. For example, it is well documented that the expansion does not converge beyond $z > 1$ and instead one should use the improved expansion parameter \cite{Cattoen:2007sk} \footnote{Note, this is an expansion in $(1-a)$, where $a$ is the scale factor normalised to $a=1$ at $z=0$ (today).}, 
\begin{equation}
\label{yparam}
y \equiv \frac{z}{1+z}. 
\end{equation}
Switching between $z$ and $y$ overcomes the problem with convergence, since higher powers of $y$, i. e. $y^n$, clearly converge. This still leaves the thorny problem concerning where best to truncate the cosmographic expansion. Related problems have been teased out elsewhere \cite{Capozziello:2008tc, Capozziello:2011tj, Capozziello:2017nbu, Capozziello:2019cav, Busti:2015xqa, Zhang:2016urt}. In this section, we focus initially on the log polynomial approach to cosmography adopted in \cite{Risaliti:2018reu, Lusso:2019akb}. We will later also briefly return to touch upon traditional cosmography. 

It is worth stressing that one must perform \textit{a Taylor expansion around $z=0$} to identify the parameters $a_i$ in terms of the single parameter $\Omega_m$ in flat $\Lambda$CDM, (\ref{a2}) - (\ref{a4}). For precisely this reason, one may worry that the above relations are only valid at low redshift. However, as is clear from the orange curve in Fig. \ref{approx_LCDM_wCDM} (a) (green in Fig.  \ref{approx_LCDM_wCDM} (b))  these relations are valid to $z \approx 4.3$ in the sense that the discrepancy with $\Lambda$CDM is still less than $1 \%$ for the canonical value $\Omega_m = 0.3$. It is worth noting that $\Omega_m = 0.3$ is a value where the log polynomial expansion performs better (recall Fig. \ref{approx_LCDM_wCDM}), so for other values, the phantom tensions we illustrate here are expected to be worse (see \cite{Banerjee:2020bjq}). For later convenience we record the $a_i$ based on canonical values: 
\begin{equation}
\label{lcdm_values}
a_2 = 2.9358, \quad a_3 = 3.54123, \quad a_4 = 1.12066. 
\end{equation}

\subsection{Mock data}     
\begin{figure}[H]
\centering
\begin{subfigure}{.5\textwidth}
  \centering
  \includegraphics[width=80mm]{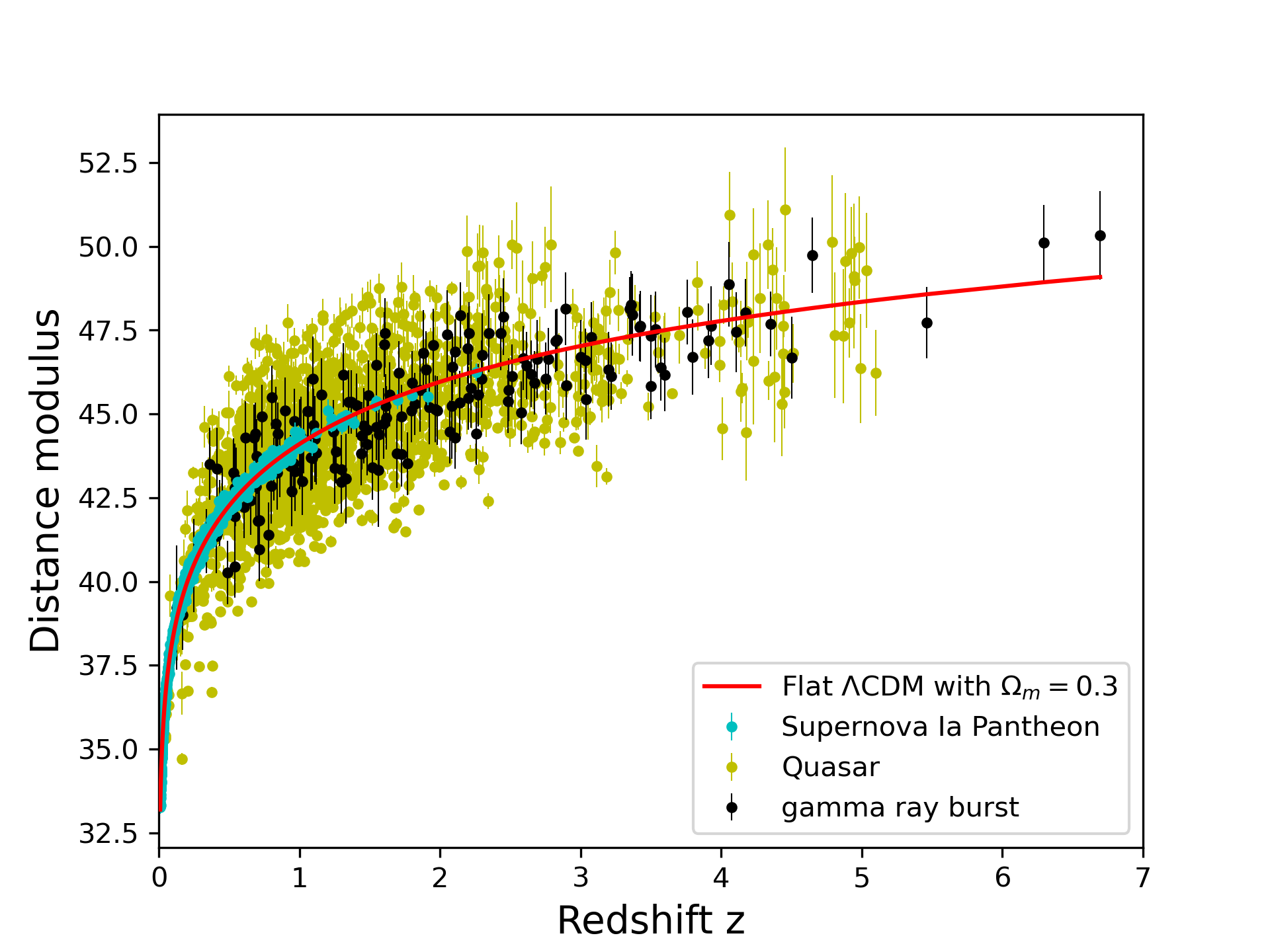}
  \caption{}
\end{subfigure}%
\begin{subfigure}{.5\textwidth}
  \centering
  \includegraphics[width=80mm]{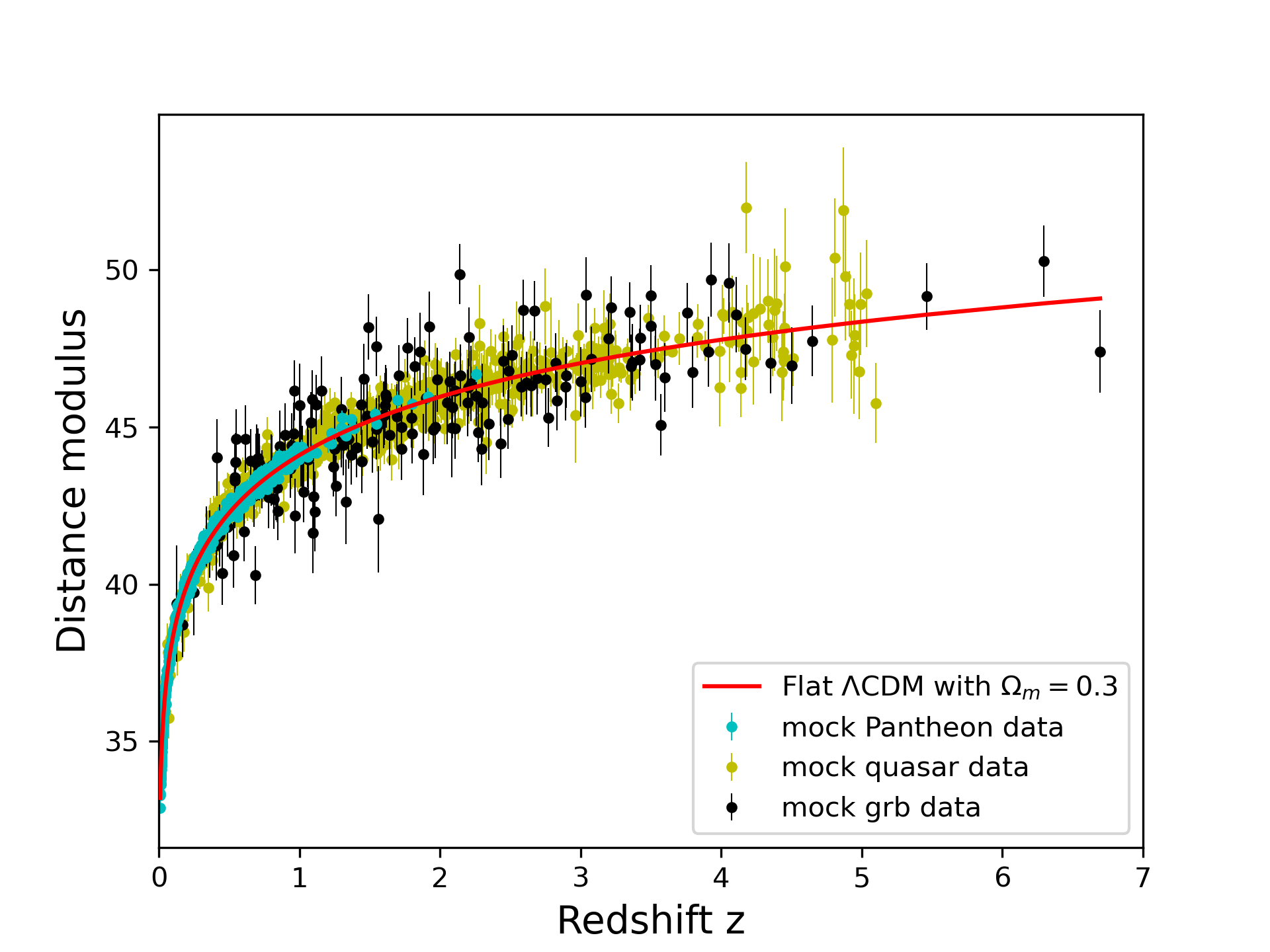}
  \caption{}
\end{subfigure}
\caption{The original Hubble diagram of SNe, QSOs and GRBs  on the left \cite{Lusso:2019akb} with mock data based on flat $\Lambda$CDM with canonical values on the right hand side.}
\label{real_mock_data}
\end{figure}

Here, we work with the original data of \cite{Lusso:2019akb} comprising SNe, QSOs and GRBs, reproduced in Fig. \ref{real_mock_data} (a). Relative to the distance moduli of the QSOs, the SNe and GRBs are displaced by $+19.36$ and $+0.54$, respectively \footnote{We thank Elisabeta Lusso and Guido Risaliti for explaining and sharing their distance modulus data. As explained in \cite{Risaliti:2015zla} (bullet points in section 2), whether one uses the flux data or the distance modulus data, it is expected that one will infer the same cosmological parameters.}. The first figure here is easy enough to understand as it is more or less,  and up to a sign, the canonical value for the absolute magnitude $M$ of Type Ia SNe. We will see later than these shifts lead to best-fit values of $H_0$ that are within $1 \, \sigma$ of the canonical value $H_0 = 70 \, \textrm{km s}^{-1} \textrm{ Mpc}^{-1}$. 

Now, let us mock up some data. For each triplet $(z_i, \mu(z_i), \Delta \mu(z_i))$, where $\mu (z_i)$ is the distance modulus and $\Delta \mu(z_i)$ denotes the error in the distance modulus of the original data, we replace the second entry with a new value $\tilde{\mu}(z_i)$ that is picked from a normal distribution around its flat $\Lambda$CDM value $\mu^{\Lambda\textrm{CDM}}(z_i)$ with standard deviation given by the error $\Delta \mu (z_i)$, i. e. $\tilde \mu(z_i) \sim \mathcal{N} (  \mu^{\Lambda\textrm{CDM}}(z_i), \Delta \mu (z_i))$. The final result is shown in Fig.  \ref{real_mock_data} (b). Compared to the real data, the mock data more closely follows flat $\Lambda$CDM, in line with expectations. Note, we have not considered the systematic uncertainties in the Pantheon SNe data, which brings it into line with QSO and GRB where all uncertainties are statistical.

\begin{figure}[H]
   \centering
   \includegraphics[width=80mm]{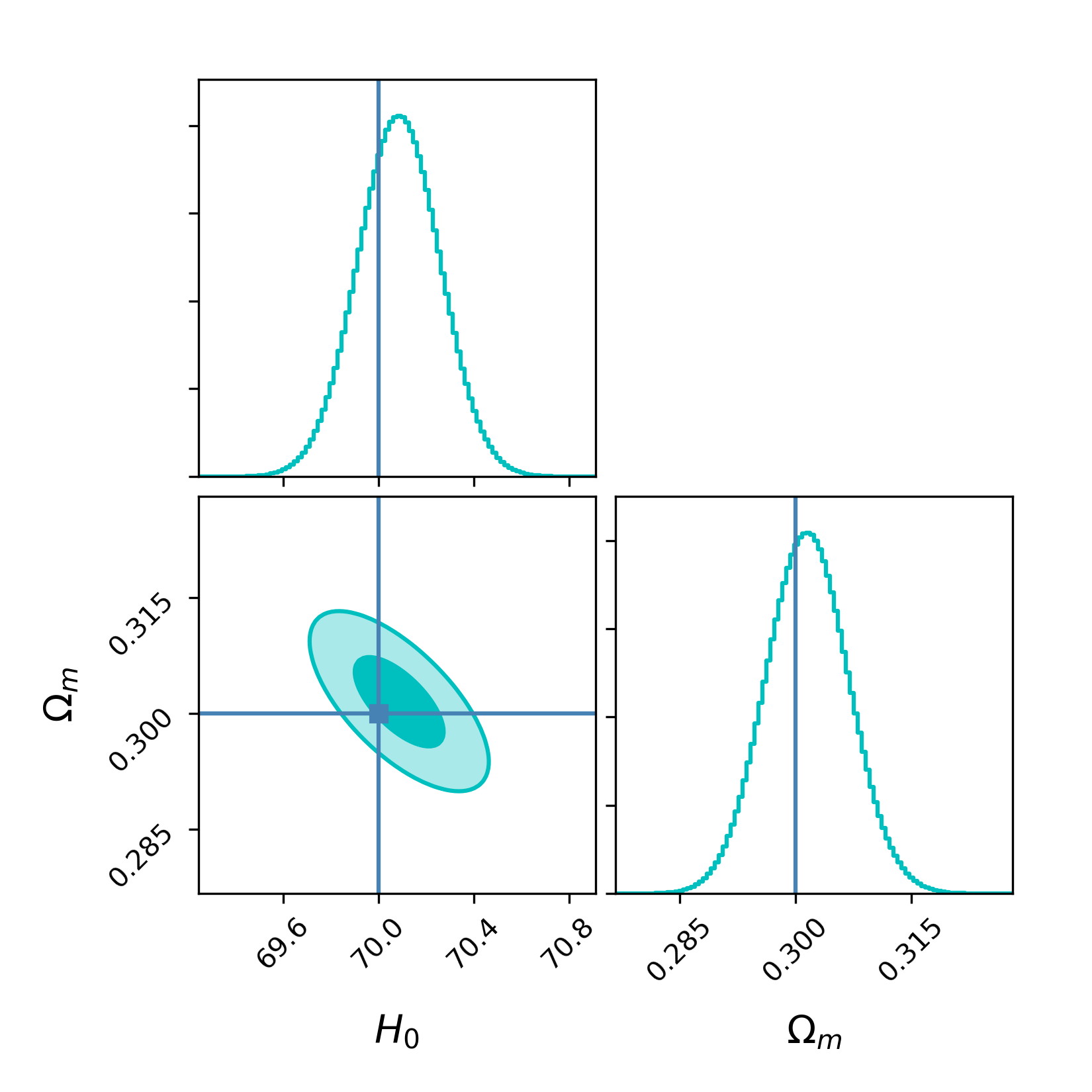}
   \caption{Best-fit values of flat $\Lambda$CDM to the mock data. The lines correspond to the canonical values $H_0 = 70$ km s$^{-1}$ Mpc$^{-1}$ and $\Omega_m = 0.3$.}
              \label{mockcheck}%
 \end{figure}

With the mock data in hand, the first thing to do is to confirm that the data is consistent with the underlying flat $\Lambda$CDM model from where it was generated. {To do so, we fit the mocked distance moduli against the flat $\Lambda$CDM model}. In Fig. \ref{mockcheck} we show the canonical values versus the best-fit values from Markov Chain Monte Carlo (MCMC) using the Python package \textit{emcee} \cite{emcee} and confirm that the returned values are within $1 \, \sigma$ of the original values. This allows us to quantify the degree of ``noise" we have added in mocking up the data. Clearly, the data is consistent with flat $\Lambda$CDM. Throughout this paper, we have checked that the number of MCMC iterations $N$ exceeds $N = 50 \, \tau$, where $\tau$ is the autocorrelation time of a given parameter. This is required to ensure \textit{emcee} chains are sufficiently converged. Moreover, we have checked that the central value of the MCMC agrees with the result of least squares fitting, so this provides a further consistency check on the best-fit values using an independent method.

\begin{figure}[H]
   \centering
   \includegraphics[width=120mm]{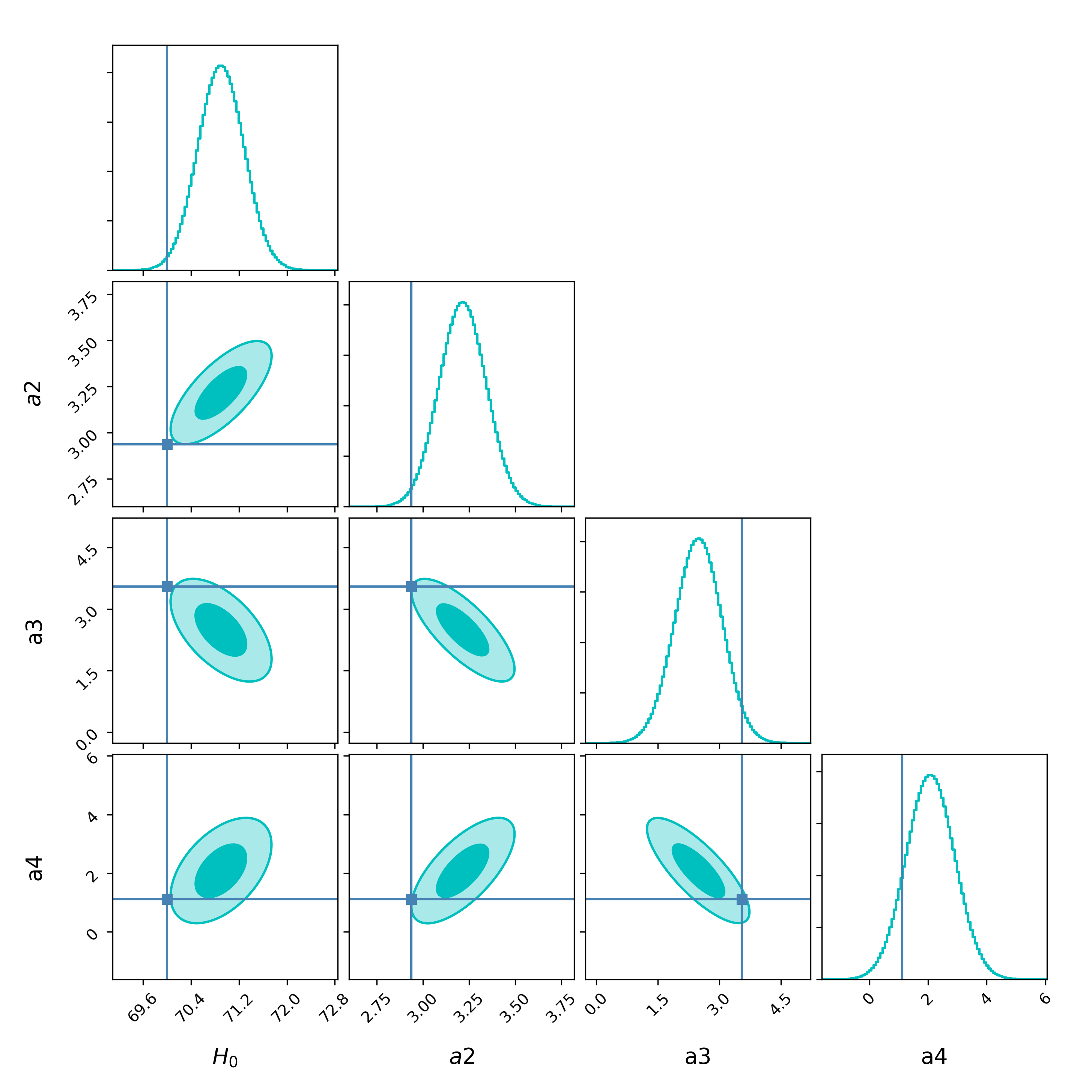}
   \caption{Best-fit values of the fourth order polynomial to the mock data. The lines correspond to the canonical values (\ref{lcdm_values}).}
              \label{mockmcmc}%
    \end{figure}

Relative to flat $\Lambda$CDM, fits of the fourth order log polynomial (\ref{dl}) make it less clear that our mock data is consistent with the standard model. Again using MCMC we identify the best-fit values, this time for the parameters $H_0, a_2, a_3$ and $a_4$. The result appears in Fig. \ref{mockmcmc}.  Here we see a noticeable difference in that the values of the parameters $H_0, a_2, a_3$ are displaced by $2 \, \sigma$ when compared to their canonical values (\ref{lcdm_values}).  $a_4$ performs marginally better, but is still outside of $1 \, \sigma$. 

Since we have replaced a two-parameter model with a four-parameter cosmographic expansion, one would expect {the fits to improve} as the number of parameters is increased. Instead, here we found the opposite. {Now, let us put this result in context. The fact that the log polynomial expansion fails to recognise flat $\Lambda$CDM mock data should be attributed to the breakdown in the expansion highlighted in section \ref{sec:scene}. The only curiosity here is that the effect is not so pronounced and the conclusion is marginal. However, this is also easy to explain. As is clear from Fig. \ref{approx_LCDM_wCDM} (a), the fourth order log polynomial through some accident approximates flat $\Lambda$CDM with $\Omega_m = 0.3$ well up to $z \sim 4$. } 

{In a companion letter \cite{Banerjee:2020bjq}, we have deliberately picked extreme examples where the approximation is so poor that the log polynomial expansion fails to recognise flat $\Lambda$CDM mock data at a statistical significance of several $\sigma$. There, we provided numerous realisations of the mock data and confirmed that the results did not change. Here we relied on one realisation, which should be sufficient when one has a large number of data points, approximately $2800$. Nevertheless, we expect any increase in tension that arises for the fourth order polynomial to be marginal, since as highlighted, through some coincidence it tracks the corresponding flat $\Lambda$CDM model with $\Omega_m = 0.3$ reasonably well. This also explains how the fourth order polynomial, despite its faults,  uncovers a \textit{bona fide} tension in the data that is not an artifact of the cosmography. As we shall show in the next section the best-fit value for flat $\Lambda$CDM to the combined dataset is in the vicinity of $\Omega_m = 0.3$} 
    
\subsection{Real GRB data}
Previously we argued that tensions, which are simply an artifact of cosmography, can arise when one fits too few parameters over too wide a range of redshifts. So far we have yet to provide a concrete example based on real data, so here we make amends. In previous studies of GRBs using traditional cosmography with the improved parameter $y$ (\ref{yparam}) deviations from flat $\Lambda$CDM have been recorded with statistical significance $2 \sim 3 \, \sigma$ \cite{Demianski:2012ra}. We will now show that such deviations may be down to the cosmographic expansion by studying the GRB dataset with the two GRBs at $z=8.1$ and $z =9.3$ reinstated. 

\begin{figure}[H]
   \centering
   \includegraphics[width=100mm]{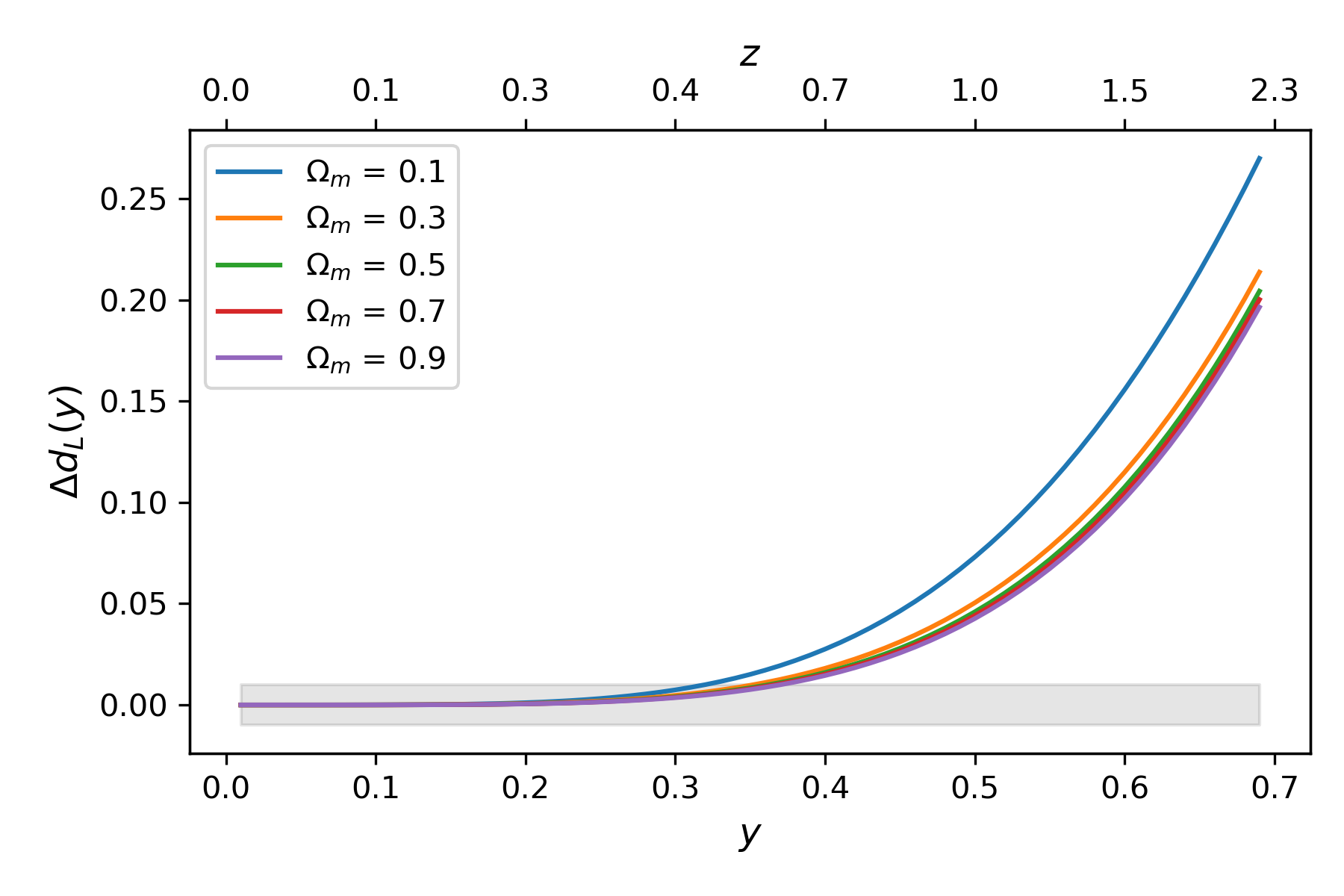}
   \caption{\% difference between the fifth order $y$-expansion and flat $\Lambda$CDM based on different values of $\Omega_m$. {We illustrate both redshift $z$ and the $y$-parameter on the horizontal axis}.}
              \label{approx2}%
    \end{figure}

Before doing so, it is once again instructive to plot the approximation inherent in the $y$-expansion based on flat $\Lambda$CDM with $(H_0, \Omega_m) = (70, 0.3)$ when it is truncated at fifth order. From Fig. \ref{approx2}, it is clear that the fifth order expansion starts to deviate from the exact analytic expression by $1$\% at $y \lesssim 0.4$ and this translates into $z \lesssim 0.67$. Clearly, beyond this redshift, comparison becomes less meaningful. {While the $y$-expansion performs well, one needs many more orders than five to approximate flat $\Lambda$CDM to $1$\% through to $z \sim 7$. Fitting such a large number of parameters, essentially makes cosmography in the $y$-parameter less practical at higher redshifts.} Interestingly, it is worth noting that Fig. 3 of \cite{Lusso:2019akb} hints at a much larger tension than $\sim 4 \, \sigma$ (potentially $\sim 8 \, \sigma$!) and this is likely due to the fact that the cosmographic expansion is impacting results. 

\begin{figure}[H]
   \centering
   \includegraphics[width=80mm]{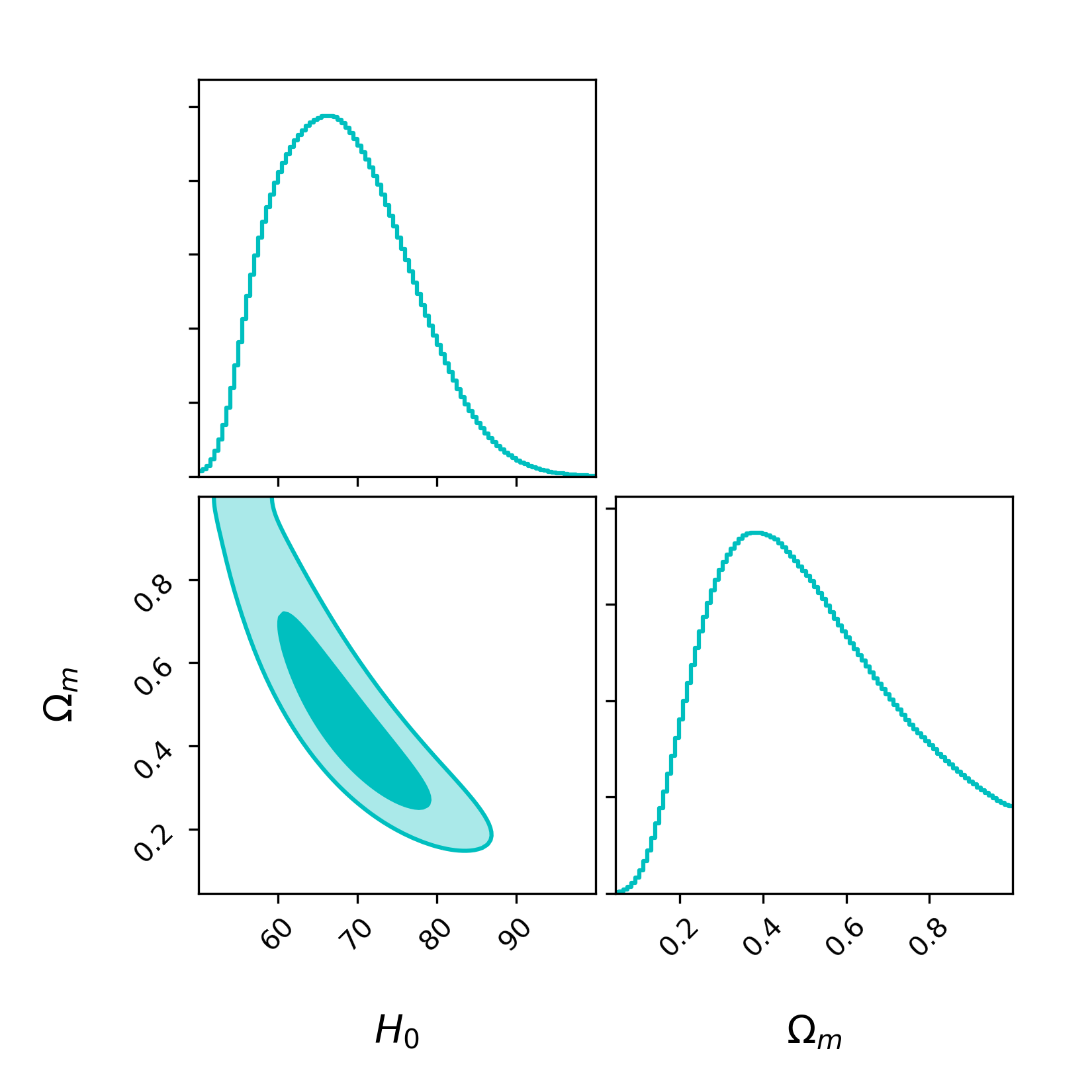}
   \caption{Best-fit values of flat $\Lambda$CDM to the GRB dataset.}%
   \label{grbplot}
    \end{figure}

Nevertheless, one can ignore this concern and blindly proceed. Let us begin by confirming that the GRB dataset on its own is consistent with flat $\Lambda$CDM. This provides us with an opportunity to test the $+0.54$ shift in the  distance moduli, which is the outcome of the calibration with respect to Type Ia SNe \cite{Lusso:2019akb}. The best-fit values are illustrated in Fig. \ref{grbplot}. Unsurprisingly, since we have few data points the errors are large, but nevertheless we see from that $H_0$ is within $1 \, \sigma$ of $H_0 = 70 \, \textrm{km s}^{-1} \textrm{ Mpc}^{-1}$. We also recognise that the best-fit value of $\Omega_m$ is just under $1\, \sigma$ removed from the canonical value. Overall, there is no hint of any deviation from the standard model. At this point we can make a timely comment. Since both the SNe and GRB data are consistent with the standard model, it is an immediate corollary that the combined dataset will also be consistent with the standard model. Nevertheless, Fig. 4 of \cite{Lusso:2019akb} reports a $2 \, \sigma$ deviation from flat $\Lambda$CDM (any value of $\Omega_m$), which is probably an artifact of the log polynomial expansion. 
    
Now, let us switch our attention to the cosmographic expansion, but here we focus on traditional cosmography, which is the setting for earlier papers, e. g.  \cite{Demianski:2012ra}. Following \cite{Zhang:2016urt} for example, we consider a power series expansion in $y$: 
\begin{equation}
d_{L}(y) = \frac{c}{H_0} \left( y + C_1 y^2 + C_2 y^3 + C_3 y^4 + C_4 y^5 \right), 
\end{equation} 
where we record the first two terms in the expansion in terms of the deceleration $q_0$ and jerk parameter $j_0$ and refer the reader to \cite{Zhang:2016urt} for the missing expressions:
\begin{equation}
C_1 = \frac{1}{2} \left(3-q_0 \right), \quad C_2 = \frac{1}{6} \left(11 - 5 q_0 + 3 q_0^2 - j_0 \right).  
\end{equation}
The canonical values for $\Lambda$CDM are $(q_0, j_0) = (-0.55, 1)$, so that the canonical flat $\Lambda$CDM values of these parameters are 
\begin{equation}
C_1 =1.775, \quad C_2 = 2.27652. 
\end{equation}
Using least squares fitting, the best-fit values of the parameters we record in Table \ref{table4} \footnote{With 162 data points least squares fitting is well suited to this task. In contrast, there is a large degeneracy in the parameters and we were unable to get the corresponding MCMC to converge. Combining the GRB dataset with SNe, we have checked that this degeneracy is broken and MCMC recovers the corresponding result based on least squares. }, where we have inferred $1\, \sigma$ confidence intervals from the returned covariance matrix and in order to save space we have only quoted a single decimal space.  The first thing to check is that the luminosity distance is indeed positive over the entire redshift range and this turns out to be the case. Next, from the best-fit values we can see that although $H_0$ is within $1 \, \sigma$ of the canonical value, the discrepancies in $q_0$ and $j_0$ are respectively $2 \, \sigma$ and $2.4 \, \sigma$. While these deviations are consistent with those reported in \cite{Demianski:2012ra} and elsewhere, since we have already checked that the data is consistent with flat $\Lambda$CDM, we confirm that these tensions are an artifact of the cosmographic expansion and not the data. 

\begin{table}[htb]            
\centering                          
\begin{tabular}{ccccc}   
\rule{0pt}{2.5ex} $H_0$ & $C_1$ & $C_2$ & $C_3$ & $C_4$ \\ 
\hline 
\rule{0pt}{2.5ex} $51.2^{+26.4}_{-26.4}$ & $-9.0^{+5.3}_{-5.3}$ & $72.2^{+28.6}_{-28.6}$ &  $-160.5^{+64.6}_{-64.6}$ & $121.3^{+50.0}_{-50.0}$ \\
\end{tabular}
\vspace{2mm}
\caption{Best-fit of traditional cosmography in the $y$ parameter to the full GRB dataset using least squares fitting.}   
\label{table4}   
\end{table}

Note, the situation with the real data closely mirrors the mock data example: consistency at the $1 \, \sigma$ level becomes a slight tension above $2 \, \sigma$. Thus, whether one works with the mock high-redshift data or real high-redshift data, we have seen in certain cases that tensions can be exacerbated by cosmographic expansions.

\end{document}